# Propagation and collisions of semi-discrete solitons in arrayed and stacked waveguides

Roy Blit and Boris A. Malomed

Department of Physical Electronics, School of Electrical Engineering, Tel Aviv University, Tel Aviv 69978, Israel

## Abstract

We consider shapes and dynamics of semi-discrete solitons (SDSs) in the known model of the set of linearly-coupled waveguides with the intrinsic cubic nonlinearity. The model applies to the description of a planar array of optical fibers, or of a stack of parallel planar waveguides, in the temporal and spatial domains, respectively, as well as to the self-attractive Bose-Einstein condensate (BEC) loaded into an array of parallel tunnel-coupled cigar-shaped traps. It was found previously that the interplay of the group-velocity dispersion, discrete diffraction (in the longitudinal and transverse directions, respectively) and intrinsic self-focusing gives rise to SDSs in the array. We here develop the variational approximation (VA) and additional analytical methods for the description of the SDSs, and study their mobility and collisions by means of systematic simulations. The VA and an exact solution of the linearized equation in the cores adjacent to the central one produce an accurate description for the family of stable fundamental onsite-centered SDS solutions, as well of surface SDSs in the semi-infinite array. The VA is also presented for transversely unstable intersite-centered solitons. In simulations, the solitons are not mobile in the discrete direction (non-soliton semi-discrete modes may be mobile across the array). Collisions between SDSs traveling in the longitudinal direction feature a threshold separating the passage and merger or destruction. The exact shape of the threshold, considered as a function of the solitons' energy, features irregularities, while its average form is explained analytically. "Shifted" collisions between SDSs centered at adjacent cores are studied too.

## 1 Introduction and the model

Waveguiding arrays offer a basis for the design of a great variety of nonlinear optical [1-9], nano-optic [10-12], and plasmonic [13-16] media, in which many types of discrete and quasi-discrete solitons can be created. Tremendous progress made in experimental and theoretical studies of these areas in the course of the last decade has been surveyed in a number of reviews [1-16]. Among other media created and investigated in this context, arrays of tunnel-coupled optical fibers [6] and nano-wires [11] enable many fundamental and applicative effects.

The most fundamental patterns predicted [17] and realized experimentally [18] in planar waveguiding arrays with the Kerr nonlinearity are discrete solitons, a well-established theoretical model which is based on the discrete nonlinear Schrödinger equation (NLSE) [19]. More recently, many theoretical results have been published for discrete solitons in similarly organized arrays of plasmonic waveguides [20].

Fiber-optic arrays may give rise to more complex modes, in the form of two-dimensional (2D) *semi-discrete solitons* (SDSs), which are built as continuous temporal solitons in the longitudinal direction (along the array), and, simultaneously, as discrete solitons in the transverse direction (across the array). Solitons of this type in planar fiber arrays had been studied theoretically some time ago, starting from Ref. [21], with expected applications to the collapse-effect compression of the solitons [22] (in the fully continuous limit, the SDSs go over into the *Townes solitons* [23] of the 2D NLSE, which are subject to the *critical collapse* in 2D) and all-optical switching and steering in the arrays [24].

Recently, settings similar to 3D fiber bundles have been created in the form of a set of parallel waveguides written by means of femtosecond pulses in bulk silica, which has made it possible to observe 3D semi-discrete spatiotemporal solitons, which are discrete in the two transverse directions [25] (see also Ref. [8] for a review). Two-component solitons of a different type, which may also be considered as semi-discrete modes, were predicted in systems with the quadratic (second-harmonic-generating) [26] and cubic [27] nonlinearities, with the continuous component propagating in a slab waveguide, and the other one carried by a discrete array attached to the slab.

The subject of the present work is the investigation of the shape, mobility, and interactions of SDSs in the model of the arrayed waveguides. The model is based on the well-known system of linearly-coupled NLSEs for envelopes $u_n(\tau)$ of the electromagnetic waves in this system [21,22,24]:

$$i\frac{\partial u_n}{\partial z} + \frac{1}{2}\frac{\partial^2 u_n}{\partial \tau^2} + \frac{1}{2}(u_{n+1} + u_{n-1} - 2u_n) + |u_n|^2 u_n = 0, \quad (1)$$

where $n$ is the number of the individual fiber (guiding core), $\tau$ the reduced temporal variable [28], and $z$ the propagation distance. In the present notation, the group-velocity dispersion (GVD) and nonlinearity coefficient, as well as the nonlinear-coupling constant accounting for the tunneling of light between adjacent waveguiding cores, are all scaled to be 1, assuming the anomalous sign of the GVD and self-focusing sign of the Kerr nonlinearity (alternatively, the normal GVD may be combined with the self-defocusing nonlinearity). The total energy of the wave field, which is a dynamical invariant of Eqs. (1), is

$$E \equiv \sum_{n=-\infty}^{+\infty} \int_{-\infty}^{+\infty} d\tau |u_n(\tau)|^2 .$$

(2) In addition to $E$, the system conserves the total momentum in the $\tau$-direction,

$i\sum_{n=-\infty}^{+\infty}\int_{-\infty}^{+\infty}d\tau\, u_n\left(\partial u_\tau^*/\partial\tau\right)$ (where the asterisk stands for the complex conjugate), and the Hamiltonian [see Eq. (35) below].

The same system of equations (1) admits an alternative realization in terms of the spatial-domain propagation, for a set of stacked planar waveguides with the self-focusing intrinsic Kerr nonlinearity. In that case, $n$ is the number of the core in the stack, $\tau$ denotes the transverse coordinate, term $(1/2)\partial^2 u_n/\partial\tau^2$ represents the paraxial diffraction in the transverse direction, and Eq. (2) defines the total power of the spatial beam.

Further, Eq. (1) may also be realized as a system of coupled Gross-Pitaevskii equations for a set of parallel cigar-shaped traps confining a self-attractive Bose-Einstein condensate (BEC), which are coupled by the tunneling of atoms across potential barriers separating the individual traps. This setting can be implemented by means of a combination of optical lattices [29], namely, a very deep one isolating the planar layer, and a moderately strong perpendicular lattice splitting the layer into the quasi-discrete set of trapping cores [30]. In terms of the BEC model, Eq. (2) defines the total number of atoms in the condensate.

In previous works [21,22,24], the SDSs in system (1) were constructed in a numerical form. Our first objective is to develop a variational approximation (VA) for them. This is a natural approach, which was elaborated in detail for fully discrete solitons in various models based on the 1D discrete NLSE [31-37], as well as for the symmetry-breaking and switching dynamics of temporal solitons in dual- [38,39] and triple- [40] core fibers. However, this method was not previously applied to SDSs. Actually, it may be quite relevant not only for the model based on Eq. (1), but also for SDSs in other systems. Results produced by the VA are reported in Section 2, and their comparison with numerical solutions is presented in Section 3, demonstrating a reasonably good agreement. The VA is applied to the onsite-centered fundamental solitons and surface solitons in the semi-infinite version of the system [41] (in terms of static state, the surface solitons are tantamount to twisted, i.e., antisymmetric, ones). For the completeness of the description, we also apply the VA to intersite-centered SDSs, which are unstable in the transverse direction (twisted solitons turn out to be unstable too, while their surface counterparts are stable in the semi-infinite system). In the limit case when almost all the energy of the SDS is trapped in the central core, the form of the components in the adjacent cores is directly found in an analytical form.

An issue of fundamental significance is the mobility of discrete solitons [19] and, accordingly, of their semi-discrete counterparts. Furthermore, in case the solitons are mobile, it is relevant to study collisions between them [32,42]. It is well known that sufficiently broad discrete solitons, if kicked, feature efficient mobility in the framework of the one-dimensional (1D) discrete NLSE with the onsite cubic nonlinearity [42], and this property may be studied by means of the VA [32,36]. Moreover, collisions between moving discrete solitons may also be analyzed with the help of this method, although in a rather cumbersome form [32].

The second major objective of the present work is to study the mobility and collisions of SDSs in the framework of Eq. (1), chiefly by means of systematic simulations. Numerical results concerning the mobility are reported in Section 3. The first conclusion is that, unlike the fully discrete solitons in the 1D discrete NLSE, the SDSs are not mobile in the discrete direction (across the array). This is explained by the fact that discrete solitons tend to be mobile when they are broad enough, which corresponds to the quasi-continuum limit [32,43], while, as mentioned above, the 2D continual counterpart of the SDS is the *Townes soliton*, that, in turn, is subject to the collapse, i.e., the catastrophic self-compression [23]. Thus, the onset of the collapse converts broad quasi-continual solitons back into the essentially discrete (narrow) ones [44]. The self-compression eventually arrests the collapse, making the SDSs strongly pinned to the underlying lattice structure, i.e., immobile. For the same reason, fully discrete 2D solitons in lattices with the cubic onsite nonlinearity demonstrate no mobility either. 2D discrete solitons are effectively mobile in settings with the saturable [45] or quadratic (second-harmonic-generating) [46] nonlinearity, where the collapse does not occur (in the continuum limit), allowing for the existence of stable broad solitons.

On the other hand, the SDSs are obviously mobile in the continual direction, which suggests the possibility to consider collisions between them. This possibility is pursued in Section 4, for both head-on collisions and those with a transverse shift between the solitons being investigated. In the former case, the collisions are quasi-elastic if the initial kick, which sets each soliton in motion, exceeds a certain critical value; below the critical value, which decreases inversely proportional to the soliton's energy, the collisions lead to *merger* of the two solitons into a single mode, or their destruction, if the solitons' energy is too small. In Section 4, the dependence of the critical kick on the energy is explained by means of an analytical estimate. A minimum value of the kick necessary to make the collision elastic is identified too for collisions between solitons with a transverse shift, whose centers are placed at adjacent cores of the array. However, in this case the collision does not cause the merger of the solitons below the critical value. Rather, the collision features inelasticity in the form of a conspicuous transfer of the energy from one soliton to the other. The latter regime is very sensitive to small changes in the initial conditions.

## 2 The variational approximation

### 2.1 Fundamental onsite-centered solitons

Stationary solutions to Eq. (1) with propagation constant $k$ are looked for as

$$u_n(\tau, z) = \exp(ikz) U_n(\tau),$$

with real functions $U_n(\tau)$ obeying the following coupled equations:

$$-kU_n + \frac{1}{2}\frac{d^2 U_n}{d\tau^2} + \frac{1}{2}(U_{n+1} + U_{n-1} - 2U_n) + U_n^3 = 0, \tag{3}$$

which can be derived from the corresponding Lagrangian,

$$L = \frac{1}{2}\sum_{n=-\infty}^{+\infty}\int_{-\infty}^{+\infty} d\tau \left[-(k+1)U_n^2 - \frac{1}{2}\left(\frac{dU_n}{d\tau}\right)^2 + U_n U_{n+1} + \frac{1}{2}U_n^4\right]. \tag{4}$$

Approximate solutions for stationary onsite-centered SDSs are sought for in the form of the following variational *ansatz*, which is a product of the usual ones adopted for discrete [31] and continual [39] solitons:

$$U_n = A\exp(-\alpha|n|)\operatorname{sech}(\eta\tau). \tag{5}$$

Here $\alpha^{-1}$, $\eta^{-1}$ and $A$ are, respectively, widths of the soliton in the discrete and continual directions, and its amplitude. The total energy of the ansatz, calculated as per Eq. (2), is $E = (2A^2/\eta)\coth\alpha$, which suggest us to eliminate the amplitude in favor of the energy:

$$A^2 = \frac{1}{2}\eta E \tanh\alpha. \tag{6}$$

The substitution of ansatz (5) into Eq. (4) and straightforward calculations lead to the following effective Lagrangian, as a function of $E$, $\eta$, and $\alpha$:

$$L_{\text{eff}} = -(k+1)E - \frac{1}{6}\eta^2 E + E\operatorname{sech}\alpha + \frac{1}{12}\eta E^2(1+\tanh^2\alpha)\tanh\alpha. \tag{7}$$

Next, the corresponding Euler-Lagrange equations, $\partial L_{\text{eff}}/\partial\alpha = 0$ and $\partial L_{\text{eff}}/\partial\eta = 0$, take the following form:

$$\sinh\alpha = \frac{1}{12}\eta E(4 - 3\operatorname{sech}^2\alpha), \tag{8}$$

$$\eta = \frac{1}{4} E \tanh \alpha \left(1 + \tanh^2 \alpha\right). \tag{9}$$

The third variational equation, $\partial L_{\text{eff}}/\partial E = 0$, yields an expression for the propagation constant,

$$k = -1 - \frac{\eta^2}{6} + \operatorname{sech}\alpha + \frac{\eta E}{6} \tanh \alpha \left(1 + \tanh^2 \alpha\right). \tag{10}$$

Using Eqs. (8) and (9), one can eliminate $\eta$ to derive an expression for $E$ in terms of $\alpha$:

$$E^2 = \frac{48 \cosh^5 \alpha}{\left(2 \cosh^2 \alpha - 1\right)\left(4 \cosh^2 \alpha - 3\right)}. \tag{11}$$

A straightforward analysis of Eq. (11) demonstrates that the soliton's energy takes values above a finite *threshold* (minimum),

$$E_{\min}^{(\text{VA})} = 4.076 \tag{12}$$

This minimum energy may be compared to the constant energy of the Townes solitons in the 2D NLSE with two continuous coordinates [23],

$$E_{\text{Townes}} \approx 5.75. \tag{13}$$

Note that the variational prediction for the Townes-soliton energy, produced by the isotropic Gaussian ansatz [47], is

$$E_{\text{Townes}}^{(\text{VA})} = 2\pi. \tag{14}$$

The global characteristic of the soliton family predicted by the VA is dependence $E(k)$, as obtained from Eqs. (9), (10) and (11), which is displayed in Fig. 1. This dependence allows one to predict the stability of the solitons by means of the well-known Vakhitov-Kolokolov (VK) criterion [48,23]: Stable may be the branch of the plot with the positive slope, $dP/dk > 0$, i.e., at $k > k_{\min} \approx 0.56$, as seen in Fig. 1. Strictly speaking, the VK criterion yields only a necessary stability criterion, but, in the present relatively simple system, it is a sufficient one too, as confirmed by numerical simulations, see below.

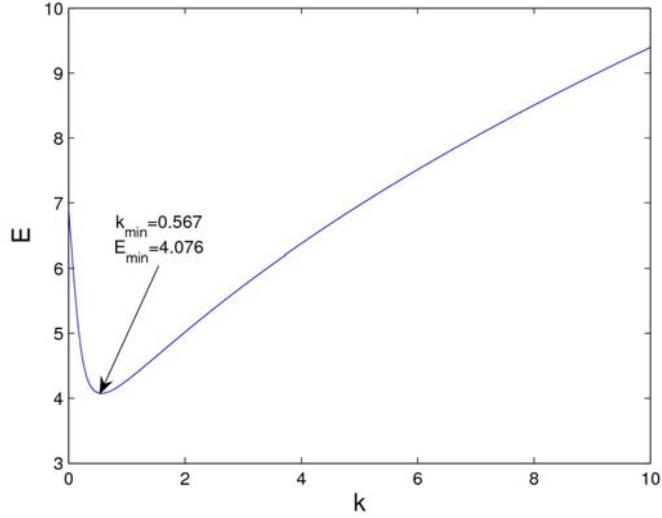

**Fig. 1:** (Color online) The energy of the fundamental onsite-centered semi-discrete soliton versus its propagation constant, as produced by the variational approximation.

An additional straightforward analysis of the variational results demonstrates that the stable branch corresponds to narrow SDSs (i.e., strongly pinned ones, which prevents their mobility in the discrete direction, see below), while the short unstable branch, predicted by the VA at $k < k_{min}$ (see Fig. 1), comprises broad solitons, which might be mobile, but the instability annuls this possibility. In particular, in the limit of $k \to 0$, a very broad SDS goes over into the Townes soliton [its energy, as given by Eq. (11), $E(\alpha = 0) = 4\sqrt{3} \approx 6.93$, is different from numerical value (13) due to the approximate form of ansatz (5), cf. the other approximate value given by Eq. (14); in fact, it is shown below that more relevant characteristics of the SDSs, such as $E_{min}$, is predicted by the VA with much better accuracy, cf. Eqs. (12) and (33)].

On the other hand, in the limit of $k \to \infty$, i.e., $E \approx 2\sqrt{2(k+1)} \to \infty$, the narrow soliton sits, chiefly, in the central core [40],

$$U_0(\tau) \approx \sqrt{2(k+1)} \operatorname{sech}\left(\sqrt{2(k+1)}\tau\right) \qquad (15)$$

with small-amplitude tails in two adjacent cores determined by the linearized version of Eq. (3) at $n = \pm 1$,

$$-(k+1)U_{\pm 1} + \frac{1}{2}\frac{d^2 U_{\pm 1}}{d\tau^2} = -\frac{1}{2}U_0. \qquad (16)$$

Substituting expression (15) into Eq. (16), it is straightforward to obtain an exact solution for the tail, valid under these assumptions:

$$U_{\pm 1}(\tau) = \tau e^{-\sqrt{2(k+1)}\tau} + \frac{1}{\sqrt{2(k+1)}}\cosh\left(\sqrt{2(k+1)}\tau\right)\ln\left(1 + e^{-2\sqrt{2(k+1)}\tau}\right) \qquad (17)$$

[in spite of its apparently asymmetric form, solution (17) is an even function of $\tau$, and the solution is exponentially localized, although it might seem divergent at $\tau \to \infty$].

To assess the degree of the self-compression of the SDS in the transverse direction, we define $R_0$ as the share of the total soliton's energy concentrated in the central core. Together with Eq. (6), ansatz (5) yields

$$R_0 \equiv E_{n=0} / E = \tanh \alpha . \qquad (18)$$

Naturally, $R_0 \to 1$ at $\alpha \to \infty$, as the soliton gets completely confined at $n=0$.

## 2.2 Twisted and surface solitons

It is well known that, in addition to the fundamental solitons, the discrete NLSE gives rise to *twisted solitons*, i.e., antisymmetric (odd) modes [49,19]. In the present system, twisted solitons can be naturally defined as those with $U_{-n}(\tau) = -U_n(\tau)$, which remain even functions of the continuous coordinate, $\tau$. In fact, the shape of the stationary twisted soliton is tantamount to that of a *surface soliton* in a semi-infinite array, which is defined as a solution to the modification of Eq. (3), with the cores existing only at $n \geq 1$:

$$-kU_n + \frac{1}{2}\frac{d^2 U_n}{d\tau^2} + \frac{1}{2}(U_{n+1} + U_{n-1} - 2U_n) + U_n^3 = 0, \text{ at } n \geq 2, \qquad (19)$$

$$-kU_1 + \frac{1}{2}\frac{d^2 U_1}{d\tau^2} + \frac{1}{2}(U_2 - 2U_1) + U_1^3 = 0, \qquad (20)$$

which can be derived from the accordingly truncated version of Lagrangian (4):

$$L = \frac{1}{2}\sum_{n=-\infty}^{+\infty}\int_{-\infty}^{+\infty} d\tau \left[-(k+1)U_n^2 - \frac{1}{2}\left(\frac{dU_n}{d\tau}\right)^2 + U_n U_{n+1} + \frac{1}{2}U_n^4\right]. \qquad (21)$$

The studies of surface solitons in various setting have recently attracted a great deal of attention [41,50-52]. In particular, the semi-infinite version of the present model gives rise to SDSs that may be considered as *surface light bullets* [41,52]

It is relevant to apply the VA to the description of the twisted/surface solitons too. To this end, the ansatz can be adopted in the form of Eq. (5) at $n \geq 1$, which implies that the soliton is attached to the edge of the array, $n=1$. Solutions with a finite distance between the soliton's peak and the edge are possible too [41], but we do not consider them here, as the respective VA would be rather cumbersome. Then, a straightforward calculation yields the corresponding energy, cf. Eq. (6), and the effective Lagrangian, cf. Eq. (7):

$$E = 2A^2 / \left[ \eta \left( e^{2\alpha} - 1 \right) \right], \tag{22}$$

$$L_{\text{eff}} = -\frac{1}{2}(k+1)E - \frac{1}{12}\eta^2 E + \frac{1}{12}\eta E^2 \tanh\alpha + \frac{1}{2}e^{-\alpha}E. \tag{23}$$

The variational equations following from this expression, $\partial L_{\text{eff}}/\partial\alpha = \partial L_{\text{eff}}/\partial\eta = \partial L_{\text{eff}}/\partial E = 0$, take the following form:

$$\eta E = 6e^{-\alpha}\cosh^2\alpha, \quad \eta = (E/2)\tanh\alpha, \tag{24}$$

$$k = -1 - \frac{\eta^2}{6} + \frac{1}{3}\eta E \tanh\alpha + e^{-\alpha}.$$

A corollary of Eq. (24) is the following expression for the energy:

$$E^2 = 12 e^{-\alpha} (\coth\alpha) \cosh^2\alpha, \tag{25}$$

cf. Eq. (11). It gives rise to a minimum value of the energy of the surface SDS,

$$(E_{\text{surf}})_{\min} \approx 3.7147, \tag{26}$$

which is attained at $\alpha = (1/2)\ln(4+\sqrt{13}) \approx 1.0144$. Accordingly, the minimum energy of the twisted soliton is twice as large, $(E_{\text{twist}})_{\min} \equiv 2(E_{\text{surf}})_{\min} \approx 7.4294$. Finally, the dependence $E(k)$, as predicted

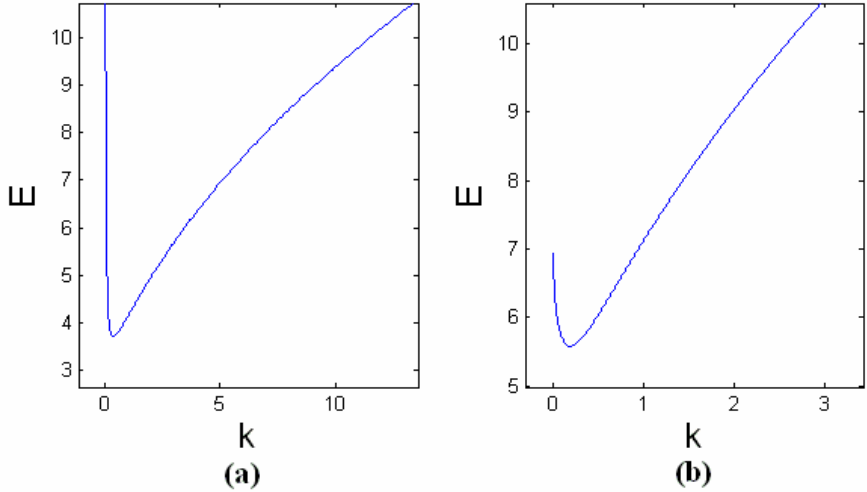

**Fig. 2:** (Color online) The energy of the surface (a) and intersite-centered (b) semi-discrete solitons versus its propagation constant, as per the variational approximation. The energy of the twisted solitons is twice that shown in (a).

## 2.3 Intersite-centered solitons

The VA can also be applied to the intersite-centered solitons (in the full lattice), which are well known as unstable stationary modes of the discrete NLSE [19].

Although the intersite SDSs are unstable too (see below), for the sake of the completeness it is relevant to briefly present the VA results for them too.

The corresponding ansatz is [cf. Eq. (5)]

$$U_n = A\exp(-\alpha(n+1))\text{sech}(\eta\tau), \text{ at } n \geq 0, \quad (27a)$$

$$U_n = A\exp(-\alpha|n|)\text{sech}(\eta\tau), \text{ at } n \leq -1, \quad (27b)$$

with total energy $E = 4A^2 / \left[\eta(e^{2\alpha} - 1)\right]$, effective Lagrangian

$$L_{\text{eff}} = -\frac{1}{2}\left(k + \frac{1}{2}\right)E - \frac{1}{12}\eta^2 E + \frac{1}{24}\eta E^2 \tanh\alpha + \frac{1}{2}e^{-\alpha}E - \frac{1}{4}e^{-2\alpha}E,$$

and the variational equations

$$\eta E = 12e^{-\alpha}\left(1 - e^{-\alpha}\right)\cosh^2\alpha, \quad \eta = (E/4)\tanh\alpha, \quad (28)$$

$$k = -\frac{1}{2} + \frac{1}{32}(E\tanh\alpha)^2 + e^{-\alpha} - \frac{1}{2}e^{-2\alpha}.$$

An equation for the energy, following from Eq. (27), is [cf. Eq. (25)]

$E^2 = 48e^{-\alpha}\left(1 - e^{-\alpha}\right)(\coth\alpha)\cosh^2\alpha$. This expression attains a minimum at $\alpha \approx 0.6183$,

$$(E_{\text{inter}})_{\min} \approx 5.576. \quad (29)$$

The fact that the threshold value (29) for the intersite SDS is larger than the minimum energy (12) for its onsite counterpart complies with the fact that the intersite-centered solitons are dynamically unstable against spontaneous transformation into the onsite states. The dependence $E(k)$ predicted by the VA for the intersite modes is displayed in Fig. 2(b).

## 3 Numerical results for single solitons

### 3.1 The numerical method

Simulations of Eq. (1) were performed by means of an advanced version of the symmetrized split-step Fourier-transform method [28] for a 2D system, with each step separated into five computational stages. The ones handling the transverse coupling were added before and after the usual stages which implemented the longitudinal-dispersion and nonlinearity terms. This method is accurate up to the fifth order in the stepsize, $O(\Delta z^5)$. In the course of the simulations, $\Delta z$ was adjusted to particular configurations, with the objective to support the necessary accuracy and computation speed. Absorbers of radiation were installed at edges of the simulation domain, in both the $\tau$- and $n$- directions.

Stationary solution of Eq. (1) for the SDSs were produced by means of the imaginary-distance-propagation algorithm, built similarly to the well-known imaginary-time algorithm [53]. To this end, the simulations were started with input

$$u_n(z=0,\tau) = A\operatorname{sech}\left(\frac{n}{\Delta n_0}\right)\operatorname{sech}\left(\frac{\tau}{\Delta \tau_0}\right), \quad (30)$$

with some amplitude $A$ and widths $\Delta n_0$, $\Delta \tau_0$, and were run forward in the imaginary distance, keeping the constant value of total energy $E$ [see Eq. (2)], until the procedure would converge to a stationary SDS mode parameterized by the given value of $E$. Iterating this process for a range of energies, we have generated a family of the stationary SDS solutions for those energies at which they exist, $E > E_{\min}$, cf. Eq. (12). If the fixed energy was too small for the existence of the SDS ($E < E_{\min}$), the simulations lead to decay (spreading out) of the pattern. The stability of the soliton family produced by this algorithm at $E > E_{\min}$ was then tested by simulations of the perturbed evolution in the real propagation distance.

Further, the mobility of the solitons and collisions between moving ones were simulated by kicking a previously found stable SDS, $u_n(\tau)$, in the discrete or continuous direction:

$$u_n^{(n-\text{kick})}(z=0,\tau) \equiv u_n(z=0,\tau)e^{ian}, \quad (31)$$

$$u_n^{(\tau-\text{kick})}(z=0,\tau) \equiv u_n(z=0,\tau)e^{-i\omega\tau}. \quad (32)$$

Note that, due to the periodicity of $\exp(ian)$, $a$ may be limited to interval $0 < a \leq \pi$. The application of the largest possible $n$-kick, $a_{\max} = \pi$, actually does not set the soliton into motion, but rather converts it into a *staggered* object [19], with $u_n = (-1)^n |u_n|$.

Due to the Galilean invariance of Eq. (1) in the $\tau$-direction, the strength of the respective kick, $-\omega$, is precisely the velocity which the kick imparts to the soliton. Collisions of SDSs moving in the continuous direction were initiated by applying opposite $\tau$-kicks to soliton pairs.

## 3.2 Families of onsite-centered semi-discrete solitons, and comparison to the analytical results

The family of stationary onsite-centered SDSs was constructed with energies exceeding a minimum value, whose numerically found value is

$$E_{\min} = 4.087, \qquad (33)$$

which is almost identical to the VA prediction given by Eq. (12). Real-$z$ simulations confirm the dynamical stability of the entire soliton family found at $E > E_{\min}$.

On the other hand, the short VK-unstable branch of the broad SDSs, predicted by the VA to the left of $E = E_{\min}$ in Fig. 1, is not produced by the numerical procedure. More specifically, the imaginary-$z$ simulations, initiated by inputs corresponding to the VA-predicted shapes of the solitons which belong to the unstable branch, quickly transform them into the stationary SDSs with the same energy, but pertaining to the stable branch of $E(k)$ (at $k > k_{\min}$ in Fig. 1). This outcome of the simulations may indeed imply the instability of the corresponding branch, rather than its nonexistence.

As an example, in Fig. 3 we present a stable soliton obtained at $E = 5.85$, which exceeds $E_{\min}$, see Eq. (33). It was generated by input configuration (30) with amplitude $A = 2.8$ and widths $\Delta n_0 = 0.32$, $\Delta \tau_0 = 0.35$. The figure shows longitudinal profiles of the SDS in five central cores of the array, i.e., at $n = 0, n = \pm 1, n = \pm 2$. The comparison with the analytically predicted profiles (15) and (17) is shown as well, for $k = 3.1$, which corresponds to the sum of the energies of the $U_0$ and $U_{\pm 1}$ components equal to the energy of the numerical solution, i.e., $E = 5.85$. In fact, the numerically found and analytically predicted profiles are indistinguishable. To characterize the degree of the self-compression of this SDS, we note that the amplitude in the central core is larger than at $n = \pm 1$ by a factor of 11.6, and the latter amplitude is 10.2 times larger than its counterpart at $n = \pm 2$.

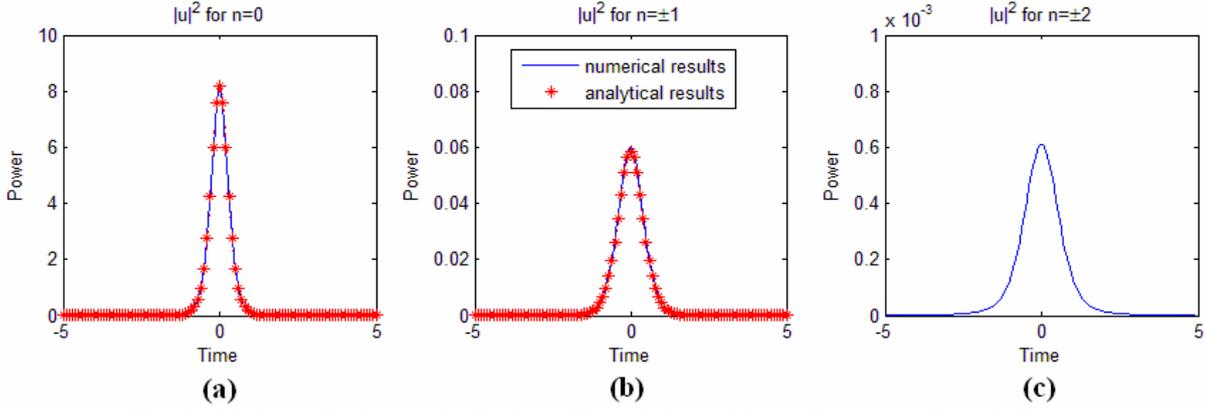

**Fig. 3**: (Color online) Longitudinal profiles of the stable onsite-centered semi-discrete soliton with energy $E$=5.85 in cores $n = 0$ (a), $n = \pm 1$ (b), and $n = \pm 2$ (c). The blue continuous line and the chain of red stars show, severally, the numerical results and analytical predictions (15) and (17).

The entire SDS family is characterized, in Fig. 4, by dependences of the amplitude, temporal width, spatial discrete width, and ratio $R_0$ [see Eq. (18)] on the total energy, $E$. The numerical widths, $\Delta n$ and $\Delta \tau$, were defined in accordance with initial ansatz (30), so that $U_1(\tau = 0)/U_0(\tau = 0) \equiv \mathrm{sech}(1/\Delta n)$, and $U_0(\tau = \Delta \tau)/U_0(\tau = 0) \equiv \mathrm{sech}(1) \approx 0.648$.

The plots also display the comparison with the results produced by the VA [see Eqs. (6), (9), (11) and (18)]. It is seen that the amplitude and temporal width, as well as $R_0$, are predicted by the VA very accurately, while the prediction of the spatial width features a discrepancy, which is explained by the fact that variational ansatz (5) was defined with the exponential form along $n$, whereas the actual shape of the soliton in the discrete direction is found to be closer to sech [for this reason, the input used in the simulations was taken in the form of Eq. (30)]. Generally, we conclude that the VA, in the combination with the additional analytical results given by Eqs. (15) and (17), produces a reasonably accurate description of the SDS family.

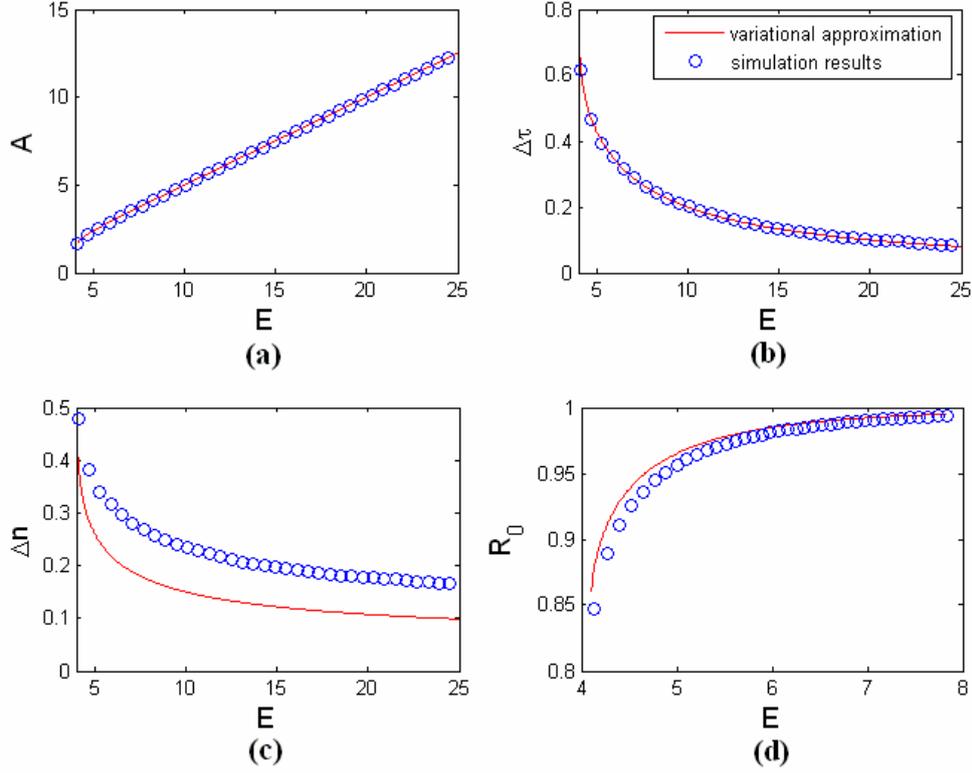

**Fig. 4**: (Color online) The amplitude (a), temporal width (b), spatial width (c), and the share of the total energy trapped in the central core (d) of the onsite semi-discrete solitons versus the energy. The blue circles and red lines show, severally, the numerical results and predictions of the variational approximation.

### 3.3 Surface solitons

The family of stationary surface SDSs attached to the edge of the semi-infinite array was found, by means if the imaginary-$z$ simulations, for energies exceeding a minimum value,

$$E_{\min}^{(surface)} = 3.78 ,$$

which is very close to the VA prediction given by Eq. (26). Real-$z$ simulations of the perturbed evolution confirm the stability of the entire surface-SDS family.

This family is characterized by dependences displayed in Fig. 5, cf. Fig. 4 for the fundamental onsite SDSs. It is seen here as well that the VA predictions are reasonably accurate.

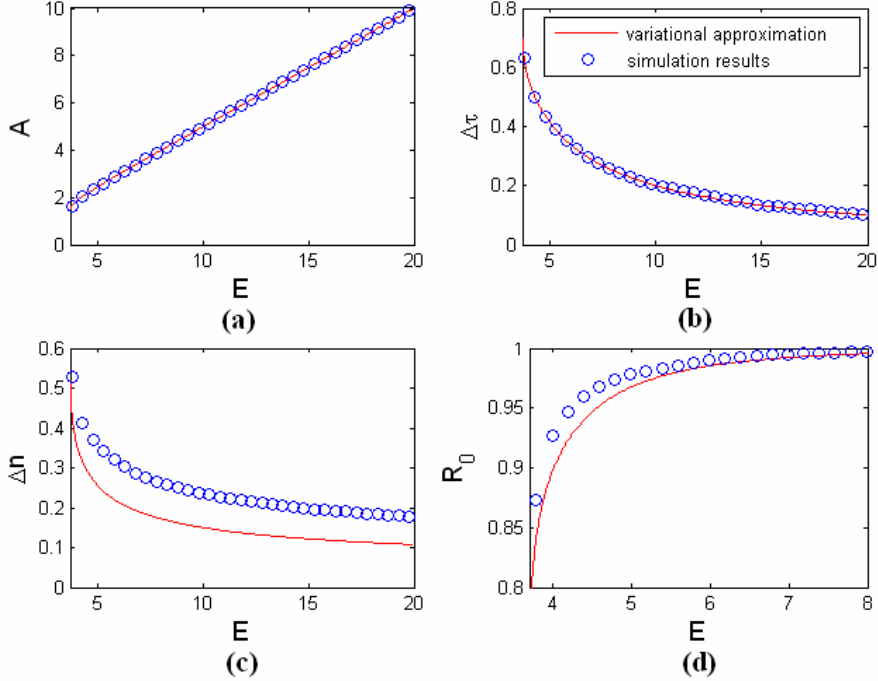

**Fig. 5**: (Color online) The amplitude (a), temporal width (b), discrete spatial width (c), and the share of the total energy trapped in the central core (d) of the semi-discrete surface solitons versus the energy. The blue circles and red lines show, severally, the numerical results and predictions of the variational approximation.

Apart from the stable onsite-centered and surface SDSs, the possibilities of the existence of families of their intersite modes [see Eq. (27)] and twisted SDSs were also investigated. In the course of the imaginary-$z$ propagation, the intersite-centered input spontaneously shifts in either direction, converting into a stable onsite soliton (as might be naturally expected, in view of the well-known instability of intersite-centered solitons in the discrete NLS equation [19]). On the other hand, the twisted input, in the form of $U_n = An \exp(-\alpha |n|) \cdot \text{sech}(\eta \tau)$, with constants $A$, $\alpha$, $\eta$, and the zero set at $n = 0$, is spontaneously transformed by the imaginary-$z$ simulations into a stable fundamental SDS centered at either $n = +1$ or $n = -1$.

## 3.4 Transverse (im)mobility of the solitons

Systematic simulations of Eq. (1) in real $z$ demonstrate that *true* stationary SDSs, produced by the imaginary-propagation-distance method, cannot be set in motion across the array by the application of the $n$-kick as per Eq. (31): The kick either initiates oscillations of the pinned soliton, or destroys it. A boundary between the different outcomes of the application of the kick has been identified in the plane of $(E, a)$, as shown in Fig. 6. The cause of the effective transverse immobility of the stable SDSs is that, in most cases, they all are quite narrow objects, being effectively localized, in the discrete direction, over ~3 cores, hence they are strongly pinned to the underlying discrete structure. On the other hand, essentially broader solitons, although being stable against small perturbations, turn out to be fragile against the application of the hard kick.

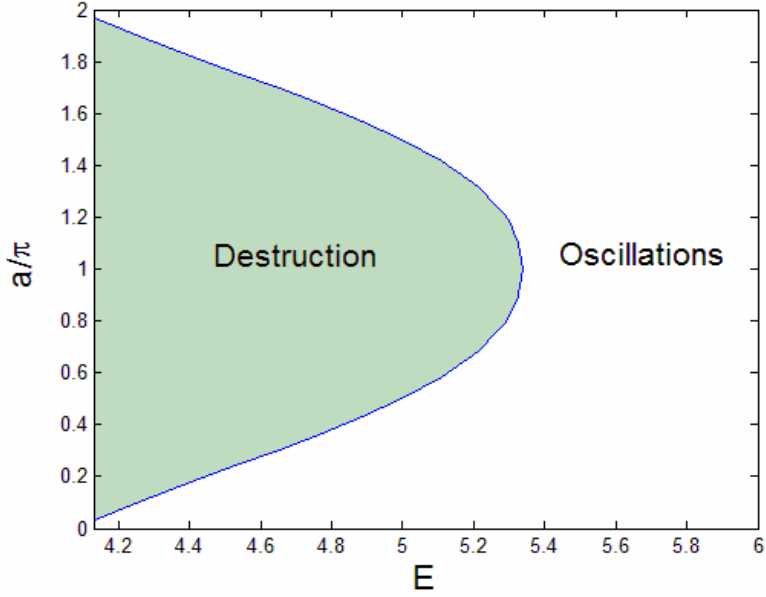

**Fig. 6**: (Color online) Regions in the plane of the soliton's energy, $E$, and strength of the $n$-kick, $a$, where the stable onsite-centered soliton reacts to the kick by oscillations or destruction. Note that the picture is periodic in the vertical direction – actually, with period $\Delta a = \pi$ (taking into regard that positive and negative values of $a$ are mutually equivalent). Values of the energy start from threshold (33). The destruction does not occur at $E > 5.3$.

On the other hand, it is possible to find non-soliton but quasi-stable broad semi-discrete localized patterns which, in the kicked form, feature effective transverse mobility, as shown in Fig. 7, for a semi-discrete pulse with energy $E = 4$. This pattern is spread over approximately 13 cores, and, in the case shown in Fig. 7, being kicked with strength $a = 1.5$, it travels distance $\Delta n = 10$ across the array, within propagation distance $\Delta z = 10$. It should be stressed that the energy of this pulse is below the threshold value $E_{min}$ of the true stationary SDSs, see Eq. (33), therefore, it eventually (but quite slowly) decays after passing a sufficiently long distance: The peak power of the field falls to a half of its initial value at $z_{1/2}(a = 1.5) = 87$. For comparison, in the case where no kick is applied to the same pattern, the half-decay distance is essentially shorter, $z_{1/2}(a = 0) = 28$, i.e., the kick effectively *stabilizes* the non-soliton mode. Because the action of the nonlinearity on such low-energy modes is weak, they can collide quasi-elastically, readily passing through each other (which was verified in simulations, but is not shown here in detail).

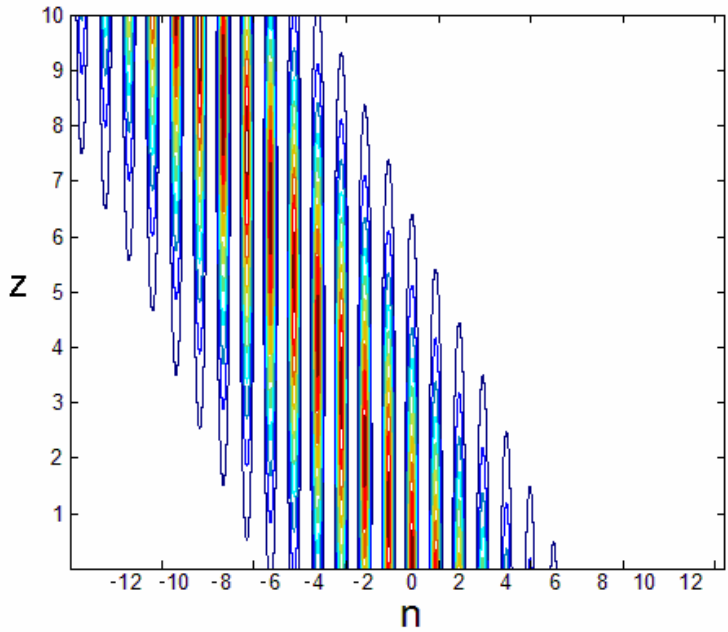

**Fig. 7**: (Color online) Quasi-stable transverse motion of an *n*-kicked non-soliton semi-discrete broad mode with energy $E = 4$, kicked with $a = 1.5$. The figure shows intensity contour plots.

A similar numerical experiment can be carried out for an initially broad semi-discrete pulse, whose total energy *exceeds* $E_{min}$. As shown in Fig. 8, the application of the transverse kick to such a "proto-soliton" sets it in motion across the array. However, the onset of the quasi-collapse causes the self-compression of the pattern, and it comes to a halt after passing a finite distance, then featuring strong oscillations in the pinned state.

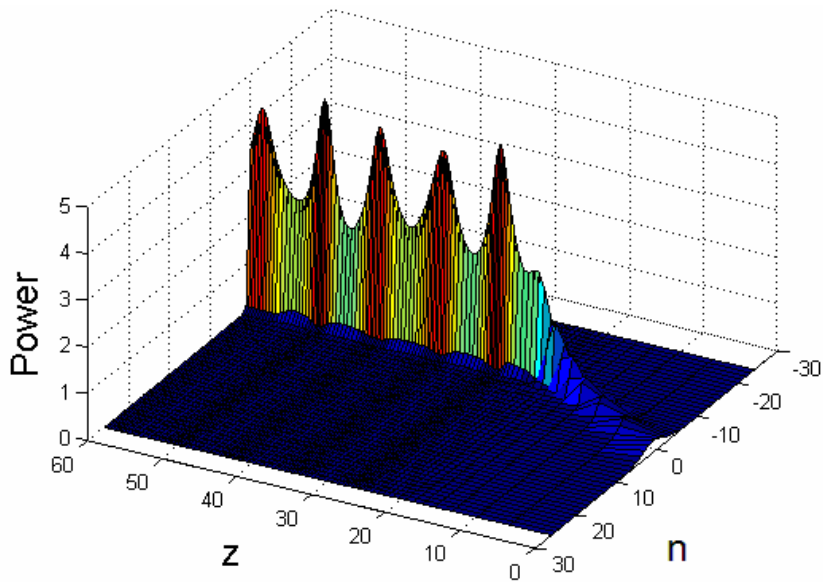

**Fig. 8**: (Color online) The evolution of an initially broad semi-discrete pulse, spread over approximately 8 cores, with energy $E = 6.67 > E_{min}$, which was kicked in the transverse direction by *a* = 1.5, see Eq. (31). The figure displays the evolution of the pulse in its central cross-section, $\tau = 0$. At first, the pulse is mobile and its center travels from *n* = 0 to *n* = -10, where it experiences the quasi-collapse and abrupt transition into the pinned mode.

## 4 Collisions between solitons in the longitudinal direction

The mobility of SDSs in the continual ($\tau$) direction is an obvious consequence of the Galilean invariance of Eq. (1) in this direction, which suggests to consider collisions between the solitons moving in opposite directions along axis $\tau$, i.e., collisions of SDSs kicked as per Eq. (32), with velocities $\pm \omega$. In addition to the fundamental interest, such collisions may have potential applications to all-optical steering of nonlinear signals in photonic devices. We here focus on the onsite-centered solitons in the full array. The issue of collisions between surface SDSs is a relevant problem too, which will be considered elsewhere.

### 4.1 Head-on collisions: The merger

The most natural case corresponds to the head-on collision between identical solitons, with equal energies *E*, moving in opposite directions along the same central core. Systematic simulations demonstrate that there is a critical value of the kick's strength, $\omega_c(E)$, such that the collision is quasi-elastic at $\omega > \omega_c(E)$, while the two solitons *merge* into a semi-discrete bound state at $\omega < \omega_c(E)$. At energies $E < 5.2$, the inelastically colliding solitons do not merge, but rather suffer mutual destruction, see below.

In fact, the very existence of the threshold, below which the collisions lead to the merger, is a noteworthy finding. Indeed, the stable SDSs are, typically, strongly localized objects, with ≥ 85% of their total energy actually concentrated in the central core, see Fig. 4(d). Therefore, one might expect that the collision between them may be almost tantamount to a collision of two solitons of the integrable NLSE which describes the central core in isolation, the collisions being completely elastic in that limit. Nevertheless, the merger, which is a typical manifestation of the nonintegrability (in particular, in the 1D discrete NLSE [32]), is readily revealed by the simulations. Naturally, $\omega_c(E)$ decreases with the increase of the energy of the colliding solitons (i.e., with the strengthening of the self-compression of the semi-discrete soliton towards the single core), as shown in Fig. 9.

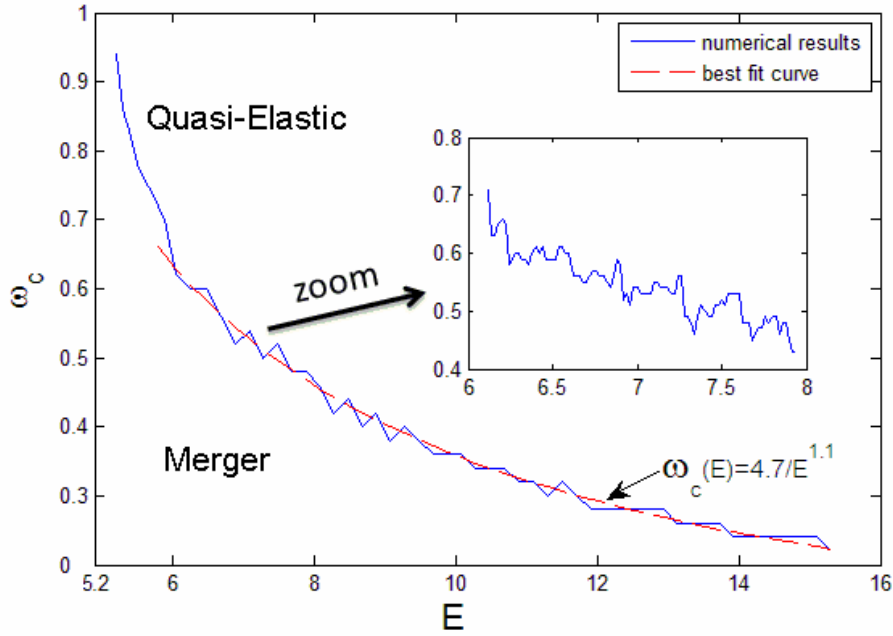

**Fig. 9**: (Color online) The broken blue line shows the critical velocity $\omega_c$, separating quasi-elastic head-on collisions between identical solitons with energy $E$, and their collision-induced merger, versus the soliton's energy, $E$, at, $E > 5.2$ (see Fig. 10 for $E_{\min} < E < 5.2$). The inset depicts this dependence with a higher resolution for $6 < E < 8$, providing a better look at irregularities in the dependence. The red dashed line is the power-law best fit to the dependence at $E > 5.8$.

We stress that irregularities in the $\omega_c(E)$ curve, clearly observed in Fig. 9, are a genuine feature of the dependence, rather than a manifestation of numerical inaccuracies (this feature is zoomed in the inset). Qualitatively similar fine-scale irregularities (which may manifest the existence of an underlying dynamical fractal structure) in the dependence of the critical velocity, separating quasi-elastic and inelastic collisions, are known in other nonintegrable systems [54].

Apart from the irregularities, the smoothed $\omega_c(E)$ dependence at $E > 5.8$ may be fitted to a power-law curve, which is found to be

$$\omega_c(E) \approx 4.7 \cdot E^{-1.1}, \tag{34}$$

i.e., the critical velocity is almost exactly inversely proportional to the solitons' energy (at $E < 5.8$, the dependence is strongly affected by transition to the regime of the collision-induced destruction of broad solitons, see below). The latter dependence can be easily explained. Indeed, the Hamiltonian of system (1) is

$$H = \frac{1}{2} \sum_{n=-\infty}^{n=+\infty} \int_{-\infty}^{+\infty} \left[ \left| \frac{\partial u_n}{\partial \tau} \right|^2 - u_n^*(u_{n+1} + u_{n-1}) - |u_n|^4 \right] d\tau. \tag{35}$$

In the case of the collision between two solitons strongly confined to the central core, where their shapes are close to that of the usual NLSE soliton, given by Eq. (15), and the weaker components in adjacent cores are approximated by Eq. (17), the main contribution to the nonintegrable part of the interaction between such solitons is produced by the first term in the Hamiltonian density on the right-hand side of Eq. (35). Taking into regard that $E \approx 2\sqrt{2k}$ for such high-amplitude narrow solitons, it is easy to estimate the largest value of the interaction term, at the point of the full overlap between the colliding solitons, which plays the role of the potential barrier impeding the passage of the solitons through each other, $H_{int} \sim 1/E$ (the estimate follows from those for the width of the soliton in the $\tau$-direction, $T \sim 1/E$, and the amplitude of the weaker component, $a \sim 1/E$). On the other hand, the net kinetic energy of the colliding solitons is $H_{kin} \approx 2E\omega^2$. Thus, the critical velocity, which is determined by equating the kinetic energy to the potential barrier, $H_{int} = H_{kin}$, is estimated as $\omega_c(E) \sim 1/E$, in agreement with the numerically generated fit (34).

As the energy of the SDS decreases, it becomes less confined in the central core, and $\omega_c$ exhibits in Fig. 9 a transition from the approximate power-law dependence (34) to a steep rise close to $E = 5.2$. Moreover, at $E < 5.2$ inelastic collisions lead to destruction of the solitons immediately after the collision, rather than their merger. As shown in Fig. 10, the critical value of the kick's strength, below which the collision results in the destruction and above which it is quasi-elastic, grows very steeply as the energy approaches $E_{min}$, the minimum energy necessary for the existence of the onsite-centered solitons [see Eq. (33)].

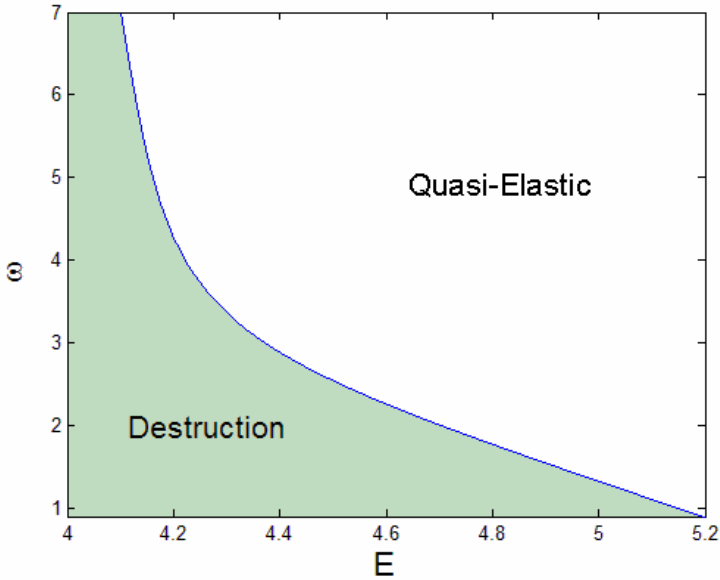

**Fig. 10**: (Color online) Regions of the quasi-elastic and destructive head-on collisions between identical solitons, in the plane of the energy, $E$, and velocities, $\pm\omega$, at $E_{\min} < E < 5.2$, where $E_{\min}$ is the existence threshold, see Eq. (33).

To analyze the merger in more detail, we will now focus on the case of $E > 5.2$, where the collisions do not cause the destruction. At velocities sufficiently close to (but smaller than) $\omega_c$, the merger proceeds through multiple collisions, which eventually give rise to a breather (an oscillating localized mode), as shown in Fig. 11. Such multiple interactions, which demonstrate repeating attempts of the solitons to separate, are a feature known in other nonintegrable systems [54]. Note also that the collision shown in Fig. 11 gives rise to a very small amount of radiative energy loss (see values of the energy indicated in the caption to the figure).

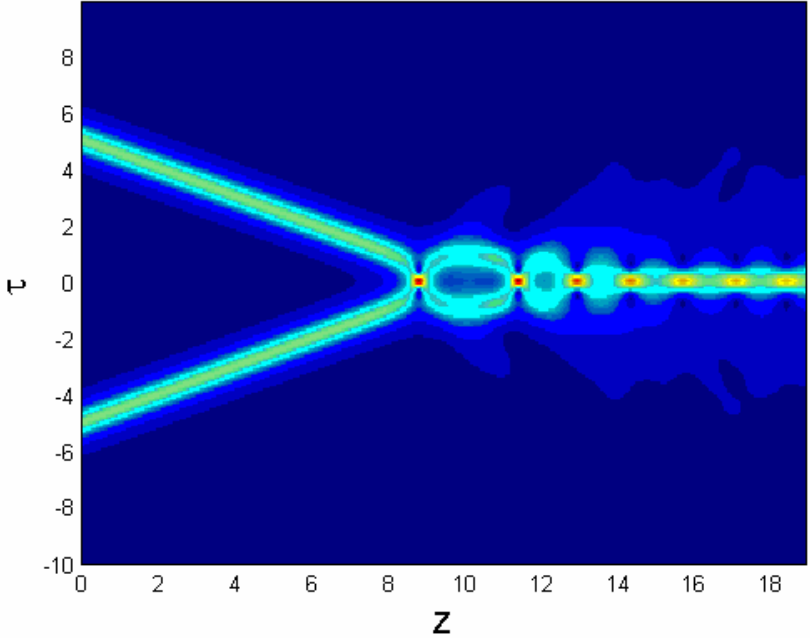

**Fig. 11**: (Color online) The merger following multiple head-on collisions of identical solitons. The contour plot of the field amplitude is shown in the central core, $n = 0$ (brighter colors correspond to higher amplitudes). Two solitons, each with energy $E = 6$, are kicked at $z = 0$ by $\omega = \pm 0.5$, and propagate until they collide at $z = 8.8$, bouncing several times and eventually merging into the breather with energy $E = 11.9$ at $z = 15$. The critical velocity in this case is $\omega_c = 0.62$.

The effect of the multiple collisions disappears at small values of the velocity. Fig. 12 displays the merger between the same solitons as in Fig. 11, but this time kicked by $\omega = \pm 0.01$. In this case, the merger occurs immediately after the first collision, while the radiative losses remain negligible.

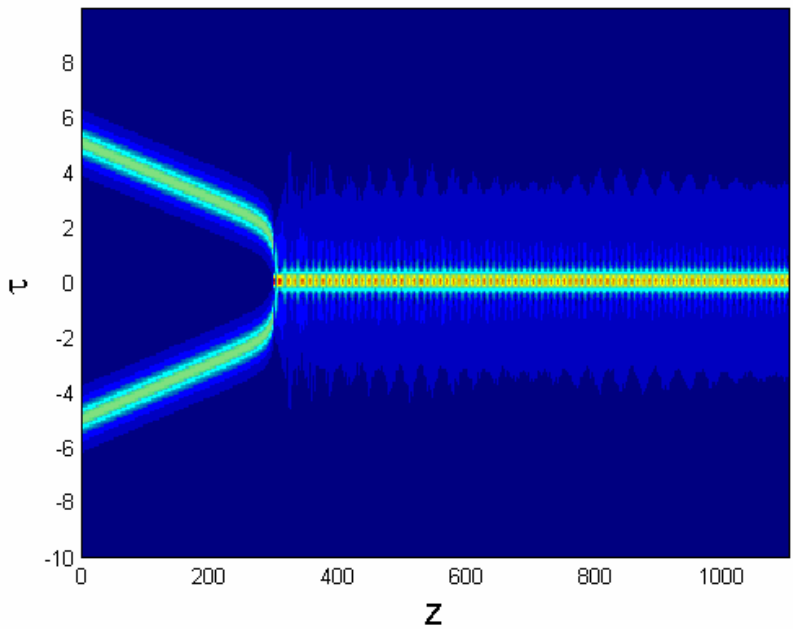

**Fig. 12**: (Color online) An example of the merger between slowly moving solitons, produced by the single collision. The solitons are the same as in Fig. 11, but kicked by $\omega = \pm 0.01$. The collision occurs at $z = 300$.

## 4.2 Quasi-elastic head-on collisions

A typical example of the quasi-elastic collision between solitons moving at velocities $\omega > \omega_c$ is shown in Fig. 13. It is seen that the collision brings about a change in velocities of the colliding solitons. In Fig. 13, the velocities are reduced by 55% after the collision. The collision-induced changes of the velocities are summarized in Fig. 14. Here too, irregularities appearing in the plots are genuine features, rather than numerical artifacts, cf. Fig. 9.

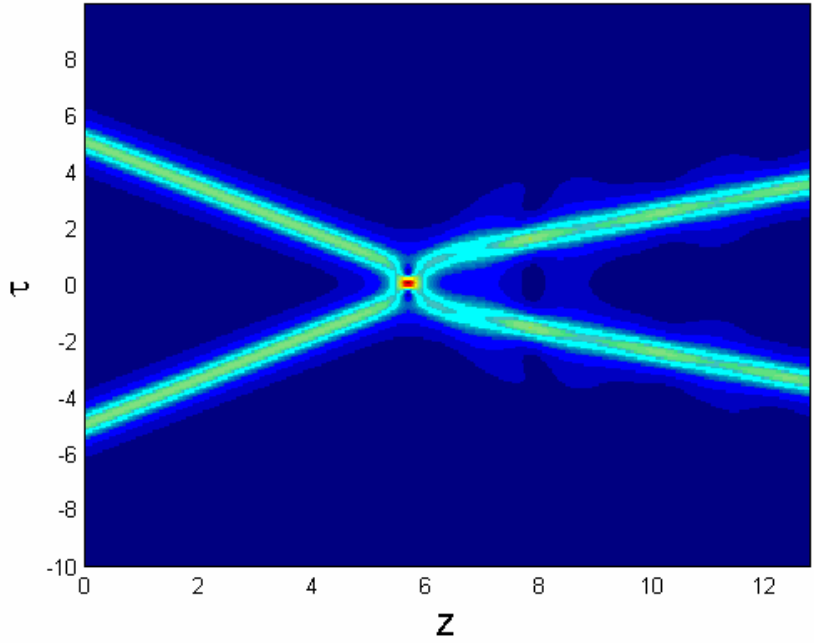

**Fig. 13**: (Color online) An example of quasi-elastic collisions. The solitons are the same as in Figs. 10 and 11, but kicked by $\omega = \pm 0.8$. They collide at $z = 5.8$ and quickly separate afterwards.

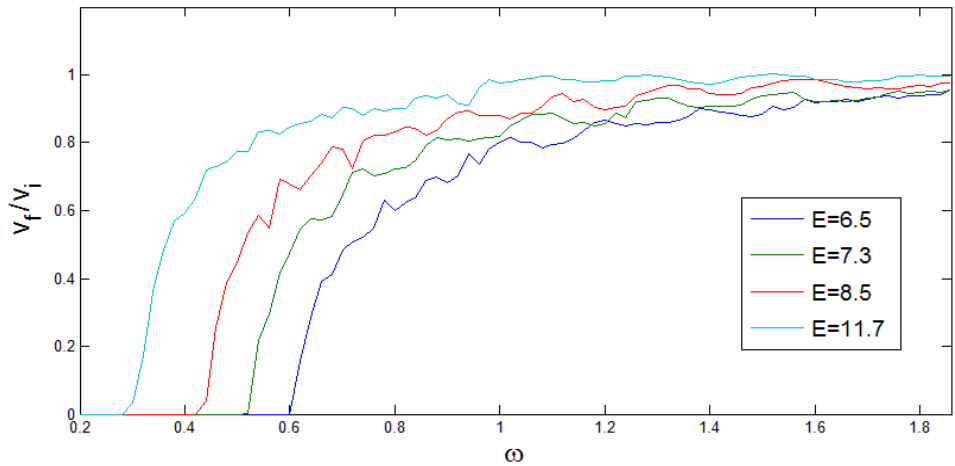

**Fig. 14**: (Color online) The collision-induced velocity change for the colliding solitons. The plots show the ratio of the final velocity to its initial value versus the kick's strength (at $\omega > \omega_c$), for different soliton energies $E$.

### 4.3 Soliton collisions with the transverse shift

Collisions between counterpropagating identical solitons centered in adjacent cores, rather than in a common one, were studied too. An example is displayed in Fig. 15, for the solitons centered at $n = 0$ and $n = 1$.

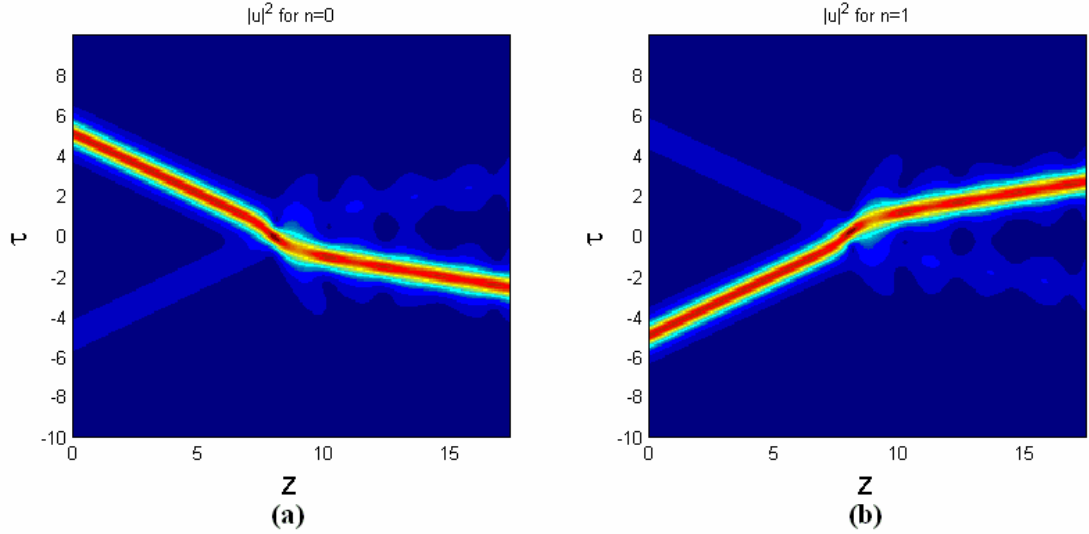

**Fig. 15**: (Color online) The collision with the transverse shift. Shown are contour plots of the field amplitude in the central cores of the two solitons, $n = 0$ (a) $n = 1$ (b). The solitons, each with energy $E = 6$, are kicked at $z = 0$ by $\omega = \pm 0.6$. The interaction causes a significant velocity change, by 62%.

As in the case of the head-on collisions, the outcome of the "shifted" collision is also characterized by a critical velocity, $\omega_c$, quasi-elastic collisions taking place at $\omega > \omega_c(E)$. However, the difference is that the interaction does not lead to merger at $\omega < \omega_c(E)$. In fact, the outcome of the interaction is not well defined in the latter case, small changes in the energy or velocity producing significantly different results. The cause of this effective instability is that, at the exact "moment" of the collision, with both solitons centered at $\tau = 0$ (in the longitudinal direction), the resulting field in the arrayed waveguide resembles that of an intersite soliton in the corresponding discrete NLSE, which is *unstable*, as discussed above (on the contrary, head-on collisions produce wave fields resembling stable onsite solitons). Below, we consider the well-defined region of quasi-elastic collisions.

The dependence of the critical velocity on the solitons' energy for the present case, produced by systematic simulations, is presented in Fig. 16. As in Figs. 9 and 14, the irregular shape of the line is a true dynamical feature, rather than a result of an insufficient numerical accuracy. From Figs. 16 and 9 we conclude that $\omega_c$ is smaller for the "shifted" collisions, in comparison with the head-on configuration. This conclusion is natural, as the interactions between two solitons centered in the same core is stronger than between solitons propagating in adjacent cores. Finally, similar to the case of the head-on interactions, the "shifted" elastic collisions give rise to a change of velocities of the involved solitons (not shown here in detail, as this effect is quite similar to that displayed in Fig. 14).

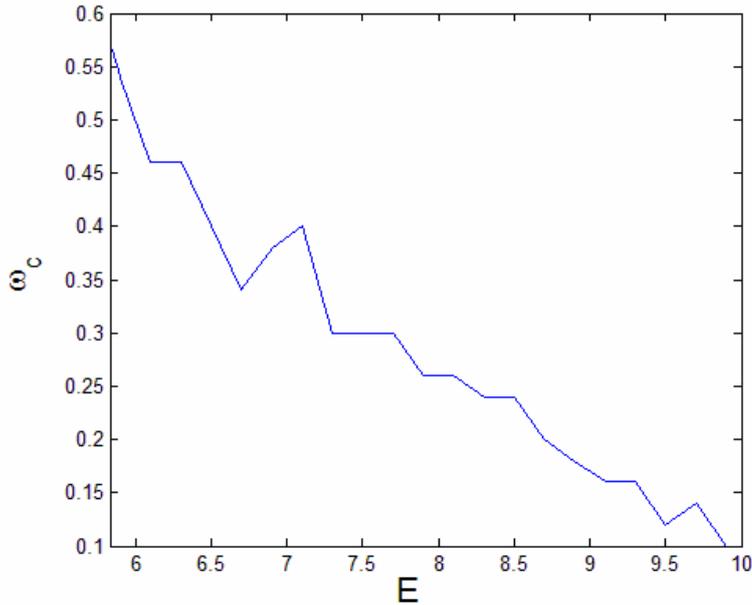

**Fig. 16**: (Color online) The critical velocity for the collisions between solitons with the transverse shift, versus the solitons' energy, at $E > E_{\min}$.

## 5 Conclusion

In this work, we have revisited the previously known model of the tunnel-coupled array of parallel waveguides, which has realizations in optics (in both the temporal and spatial domains), as well as in BEC. It was known [21,22,24,40] that the arrest of the 2D collapse by the transverse discreteness gives rise to SDSs (semi-discrete solitons) in this system. In the present work, we aimed to develop the VA (variational approximation) and other analytical methods for the description of the SDSs, and to study, by means of systematic simulations, their mobility and collisions. It has been found that the VA yields quite accurate results for the families of stable onsite-centered and surface-mode SDSs. The exact solution of the linearized equations for the cores adjacent to the central one [see Eqs. (15) and (17)] provides a good approximation for very narrow solitons too. The simulations have shown that the true solitons have no mobility across the array (non-soliton semi-discrete patterns may be efficiently mobile in the transverse direction, and may actually be temporarily stabilized by the corresponding kick). The systematic analysis of the collisions between the onsite-centered SDSs moving in the continual direction clearly shows a sharp boundary between quasi-elastic interactions and merger. The exact form of the boundary contains irregular features, but its average shape can be explained by means of an analytical estimate. The very fact of the merger caused by the collision between the SDSs, which are strongly localized in the transverse direction, is a noteworthy finding, as, in the limit when almost all the energy is trapped in the central core, one might expect completely elastic collisions governed by the usual NLSE. Collisions between solitons centered at two adjacent cores, rather than in the same one, have been considered too.

As the next step of the analysis, it is possible to look for higher-order modes, built as *N*-solitons (actually, breathers), produced by multiplying the stationary SDS by integer factor *N*. Such exact solutions of the single-core NLSE are well-known objects [28]. We have checked that SDSs for $N = 2$ are robust breathing modes, which are similar to their counterparts in the single-core NLSE. It remains to study interactions between them in the framework of the present system.

A more challenging generalization may be a semi-discrete array in the form of a fiber bundle with the structure of a discrete 2D lattice in the transverse plane, similar to the one which was recently used to create the semi-discrete spatiotemporal solitons [25]. In particular, it would be interesting to study the corresponding SDSs with the structure of discrete vortices [55,19] in the transverse lattice.